\documentclass[]{emulateapj}

\let\lsim=\la
\let\gsim=\ga

\slugcomment{Accepted for publication in the Astrophysical Journal}
\shorttitle{Obscured Star Formation in GRB Host Galaxies}
\shortauthors{B.~Hatsukade et al.}

\begin{document}

\title{Constraints on Obscured Star Formation in Host Galaxies of Gamma-ray Bursts}

\author{Bunyo~Hatsukade\altaffilmark{1}, 
		Tetsuya~Hashimoto\altaffilmark{2},
 		Kouji~Ohta\altaffilmark{1}, 
 		Kouichiro~Nakanishi\altaffilmark{2,3,4}, 
 		Yoichi~Tamura\altaffilmark{5}, 
 		and Kotaro~Kohno\altaffilmark{5,6}
 		}

\altaffiltext{1}{Department of Astronomy, Kyoto University, Kyoto 606-8502}\email{hatsukade@kusastro.kyoto-u.ac.jp}
\altaffiltext{2}{National Astronomical Observatory of Japan, 2-21-1 Osawa, Mitaka, Tokyo 181-8588}
\altaffiltext{3}{Joint ALMA Office, Alonso de C\'{o}rdova 3107, Vitacura, Santiago 763 0355, Chile}
\altaffiltext{4}{The Graduate University for Advanced Studies (Sokendai), 2-21-1 Osawa, Mitaka, Tokyo 181-8588, Japan}
\altaffiltext{5}{Institute of Astronomy, the University of Tokyo, 2-21-1 Osawa, Mitaka, Tokyo 181-0015}
\altaffiltext{6}{Research Center for the Early Universe, University of Tokyo, 7-3-1 Hongo, Bunkyo, Tokyo 113-0033, Japan}

\begin{abstract}
We present the results of the 16-cm-waveband continuum observations of four host galaxies of gamma-ray bursts (GRBs) 990705, 021211, 041006, and 051022 using the Australia Telescope Compact Array. 
Radio emission was not detected in any of the host galaxies. 
The 2$\sigma$ upper limits on star-formation rates derived from the radio observations of the host galaxies are 23, 45, 27, and 26~$M_{\odot}$~yr$^{-1}$, respectively, which are less than about 10 times those derived from UV/optical observations, suggesting that they have no significant dust-obscured star formation. 
GRBs 021211 and 051022 are known as the so-called ``dark GRBs'' and our results imply that dark GRBs do not always occur in galaxies enshrouded by dust. 
Because large dust extinction was not observed in the afterglow of GRB~021211, our result suggests the possibility that the cause of the dark GRB is the intrinsic faintness of the optical afterglow. 
On the other hand, by considering the high column density observed in the afterglow of GRB~051022, the likely cause of the dark GRB is the dust extinction in the line of sight of the GRB. 
\end{abstract}


\section{Introduction}

One of the key issues in astronomy is to understand when and how stars were formed in the history of the universe. 
The cosmic star-formation history has been explored from local to high-redshift universe \citep[e.g.,][]{mada96, lill96}. 
Long-duration gamma-ray bursts (GRBs) are thought to be powerful indicators of star-formation activity in the distant universe \citep[e.g.,][]{tota97} because 
(1) they are detectable even at $z > 8$ \citep{tanv09, salv09} 
and 
(2) they are considered to occur as a result of the deaths of massive stars \citep[e.g.,][]{stan03} and are closely associated with star formation in host galaxies. 
Host galaxies of GRBs have been studied mainly by UV, optical, and near-infrared (NIR) observations. 
The majority of GRB host galaxies are blue, subluminous, low-mass star-forming galaxies \citep[e.g.,][]{fynb03, fruc06} with star-formation rates (SFRs) of $\sim$0.1--10~$M_{\odot}$~yr$^{-1}$ \citep[e.g.,][]{sava09, sven10}. 
On the other hand, massive star-forming host galaxies have also been discovered \citep[e.g.,][]{hash10, sven11}. 
Observations at the mid-IR, submillimeter, and radio ranges show that some GRB host galaxies have a large amount of star formation obscured by dust \citep[e.g.,][]{berg03, lefl06}. 
It is found that star-formation activity at $z \gsim 1$ is dominated by IR-luminous dusty galaxies \citep[e.g.,][]{lefl05, take05}. 
If GRBs trace star-forming activity, they could occur in dust-obscured regions \citep[e.g.,][]{djor01}.

Dust extinction is a possible cause of mysterious events of so-called ``dark GRBs'' \citep[e.g.,][]{fynb01, perl09}. 
Dark GRBs, which are defined by the faintness of the optical afterglow compared with that expected from X-ray afterglow \citep{jako04}, make up about 25--40\% of GRBs \citep{fynb09, grei11}. 
The possible explanations for this are that 
(1) optical afterglows are obscured by dust within host galaxies distributed globally or in the vicinity of GRBs; 
(2) optical afterglow is intrinsically faint owing to low-luminous GRB events, fast-declining afterglows, or the low-density of the ambient medium; 
and (3) GRBs are at high redshift and the optical afterglows are absorbed by intervening neutral hydrogen. 
Detailed studies on GRB host galaxies have been mostly conducted by using UV/optical observations, which are biased toward less dusty galaxies. 
Recent extensive follow-up campaigns of GRBs and their host galaxies suggest that a large fraction of GRBs occur in dusty environments \citep[e.g.,][]{perl09, grei11}. 
In order to obtain a clear understanding of GRB host galaxies, it is essential to study them using methods that are unaffected by dust extinction. 
Radio observations are effective measures for examining star-forming activity without the effect of dust obscuration. 
So far, only a small fraction of GRB host galaxies have been observed at radio wavelength \citep[e.g.,][]{berg03, stan10}.

In this paper, we report 16-cm (2-GHz) waveband continuum observations for four GRB host galaxies. 
These results were obtained by using the Australia Telescope Compact Array (ATCA). 
Section~\ref{sec:targets} describes the target GRB host galaxies. 
Section~\ref{sec:observation} outlines the observations, data reduction, and results. 
In Section~\ref{sec:discussion}, the constraints on SFRs are derived and obscured star formation in the GRB host galaxies is discussed as well as the cause of dark GRBs. 
In Section~\ref{sec:summary}, a summary of the findings is presented. 
Throughout the paper, we adopt a cosmology with $H_0=70$ km~s$^{-1}$ Mpc$^{-1}$, $\Omega_{\rm{M}}=0.3$, and $\Omega_{\Lambda}=0.7$.

\section{Targets}\label{sec:targets}

We observed the host galaxies of GRBs~990705, 021211, 041006, and 051022. 
In order to obtain tight constraints on the obscured star formation, we selected the targets at $z \le 1$ with SFRs well derived from UV/optical observations. 
The UV/optical SFRs, as hereafter stated, are corrected for dust extinction.

The GRB~990705 host galaxy is a star-forming galaxy at $z = 0.8424$ \citep{lefl02}. 
The SFR derived from UV luminosity is SFR(UV) $\sim$5--8 \citep{lefl02} and 3.31~$M_{\odot}$~yr$^{-1}$ \citep{sven10}. 
The SFR based on the [O\,{\sc ii}] line is SFR([O\,{\sc ii}]) = 6.96~$M_{\odot}$~yr$^{-1}$ \citep{sava09}. 
The galaxy is detected with {\sl Spitzer} at 4.5 and 24~$\mu$m \citep{lefl06}. 
The IR luminosity inferred from the 24-$\mu$m detection is $1.8^{+2.1}_{-0.6} \times 10^{11}$~$L_{\odot}$, classifying the galaxy as a luminous IR galaxy (LIRG). 
The SFR derived from the IR luminosity is SFR(IR) = $32^{+37}_{-11}$~$M_{\odot}$~yr$^{-1}$. 
\cite{cast06} derived a SFR of 4.5--173~$M_{\odot}$~yr$^{-1}$ based on the optical--radio spectral energy distribution (SED) that included the {\sl Spitzer} photometry. 
The SFR derived from the IR observations is higher than the extinction-corrected SFR derived from UV/optical observations shown above, suggesting the existence of dust-obscured star formation, which cannot be traced at UV/optical wavelengths. 
\cite{berg03} failed to detect significant radio continuum emission at 1.39~GHz ($S_{\rm 1.39~GHz} = 52 \pm 32$~$\mu$Jy, which corresponds to a SFR = $190 \pm 165$~$M_{\odot}$~yr$^{-1}$).

GRB~021211 at $z = 1.006$ \citep{vree03} is classified as a dark GRB owing to its faintness of optical afterglow \citep{fox03, li03, crew03}. 
\cite{dell03} found supernova signatures in the spectrum of the optical afterglow. 
The UV-derived SFR is SFR(UV) = 3.01~$M_{\odot}$~yr$^{-1}$ \citep{sava09} and 6.95~$M_{\odot}$~yr$^{-1}$ \citep{sven10}.

The GRB~041006 associated with a supernova occurred at $z  = 0.716$ \citep{fuga04, stan05, sode06}. 
The SFRs of the host galaxy is 
SFR([O\,{\sc ii}]) = 0.34~$M_{\odot}$~yr$^{-1}$ \citep{sava09} and SFR(UV) = 1.17~$M_{\odot}$~yr$^{-1}$ \citep{sven10}.

GRB~051022 is another dark GRB whose optical/NIR afterglow was undetected. 
Millimeter and radio interferometer observations detected the afterglow and pinpointed the position of the host galaxy at $z = 0.807$ \citep{cast07, rol07}. 
The host galaxy is a luminous blue compact star-forming galaxy with a rest-frame $B$-band magnitude of $M_B = -21.8$~mag \citep{cast07}. 
The optical/NIR afterglow was not detected despite a bright X-ray afterglow and a high absorption column density. 
The X-ray spectrum of the afterglow shows a strong absorption corresponding to $N_{\rm H} = 3$--$9 \times 10^{22}$~cm$^{-2}$ \citep{naka06, cast07}. 
The galaxy shows active star formation with SFR([O\,{\sc ii}]) = 34.64~$M_{\odot}$~yr$^{-1}$ \citep{sava09} and SFR(UV) = 23.85~$M_{\odot}$~yr$^{-1}$ \citep{sven10}.

\section{Observations and Results}\label{sec:observation}

The ATCA observations were conducted on September 14 and 15, 2011 using 6 antennas in the 6B array configuration (baseline lengths of 214--5969~m). 
The Compact Array Broadband Backend (CABB) was used in the CFB 1M-0.5k mode with $2048 \times 1$-MHz channels centered at 2100~MHz (16-cm band). 
PKS~1934$-$638 was observed at the beginning of the observations as a flux and bandpass calibrator. 
PKS~0252$-$712, PKS~0741$-$063, PKS~0106+013, and PKS~0019$-$000 were observed as phase calibrators for the host galaxies of GRBs~990705, 021211, 041006, and 051022, respectively. 
To fill in the \textit{uv} plane as much as possible, we performed cyclic observations of the targets with a short integration time (15--30 minutes).

Data reduction and imaging were carried out using the MIRIAD package \citep{saul95}. 
The 16-cm band is severely affected by radio frequency interference, and hence, we carefully flagged the data. 
Because the CABB data cover a large fractional bandwidth at the 16-cm band, the calibrations were performed by using a multi-frequency strategy to account for the amplitude dependence on frequency. 
The calibrated data were imaged with using multi-frequency synthesis. 
Uniform weighting was adopted to minimise the sidelobe levels. 
The image was deconvolved with the task MFCLEAN.

We did not detect radio emission from any of the host galaxies. 
The rms noise levels are 8.7, 11, 15, 11~$\mu$Jy~beam$^{-1}$ for GRBs~90705, 021211, 041006, and 051022, respectively, which are measured at an area near the central position without any source or significant sidelobes. 
In the following section, we discuss the constraints using the 2$\sigma$ upper limits.

\begin{table}
\begin{center}
\caption{Results of the ATCA 16~cm Observations \label{tab:results}}
\begin{tabular}{cccc}
\tableline\tableline
GRB & $z$ & rms noise & SFR \\
    &     & ($\mu$Jy~beam$^{-1}$) & ($M_{\odot}$~yr$^{-1}$) \\
\tableline
990705 & 0.842 & 8.65 & $<$23 \\
021211 & 1.006 & 11.4 & $<$45 \\
041006 & 0.712 & 15.0 & $<$27 \\
051022 & 0.807 & 11.0 & $<$26 \\
\tableline
\end{tabular}
\tablecomments{Upper limits are 2$\sigma$.}
\end{center}
\end{table}

\section{Discussion}\label{sec:discussion}

\subsection{Constraints on Star-Formation Rate}

We derive the 2$\sigma$ upper limits on the SFRs of the host galaxies using the rms noise levels. 
We use the relation between radio flux and the SFR for starburst galaxies derived by \cite{yun02} as follows: 
\begin{eqnarray}
S(\nu_{\rm obs}) 
&=& \Bigg\{ 25 f_{\rm nth} \nu_0^{-\alpha} + 0.71 \nu_0^{-0.1} \nonumber \\
& & +1.3 \times 10^{-6} \frac{\nu_0^3\left[1 - e^{-(\nu_0/2000)^{\beta}}\right]}{e^{0.048\nu_0/T_d}-1} \Bigg\} \nonumber \\
& &\times \frac{(1+z){\rm SFR}}{D_L^2},
\end{eqnarray}
where $S(\nu_{\rm obs})$ is the observed flux density in Jy, $\nu_0$ is the rest frequency in GHz, $f_{\rm nth}$ is a scaling factor, $\alpha$ is the synchrotron spectral index, $\beta$ is the dust emissivity, $T_d$ is the dust temperature in K, $D_L$ is the luminosity distance in Mpc, and SFR is in $M_{\odot}$~yr$^{-1}$. 
The 1st, 2nd, and 3rd term on the right-hand side represent the contribution of non-thermal synchrotron emission, free-free emission, and thermal dust emission, respectively. 
We adopt $\alpha = 0.6$, $\beta = 1.35$, $T_d = 58$~K, and $f_{\rm nth} = 1$, which are selected for IR-selected starburst galaxies \citep{yun02} and were used in previous radio studies on GRB host galaxies \citep{berg03, stan10}. 
The 1st non-thermal term dominates the flux density at the observed frequency of 2.1~GHz. 
The 2$\sigma$ upper limits on SFRs derived using the equation (1) are 23, 45, 27, and 26~$M_{\odot}$~yr$^{-1}$ for the host galaxies of GRBs~990705, 021211, 041006, and 051022, respectively (Table~1). 
The upper limits on SFRs are less than about 10 times the UV/optical SFRs (Section~\ref{sec:targets}), suggesting no significant dust-obscured star formation, which is undetectable at UV/optical wavelengths. 
Although the {\sl Spitzer} observations indicate that the GRB~990705 host galaxy has obscured star-forming activity \citep{lefl06}, our 2$\sigma$ upper limit on SFR is consistent with the SFRs derived from the IR observations after considering their uncertainties.

\begin{figure}
\includegraphics[width=\linewidth]{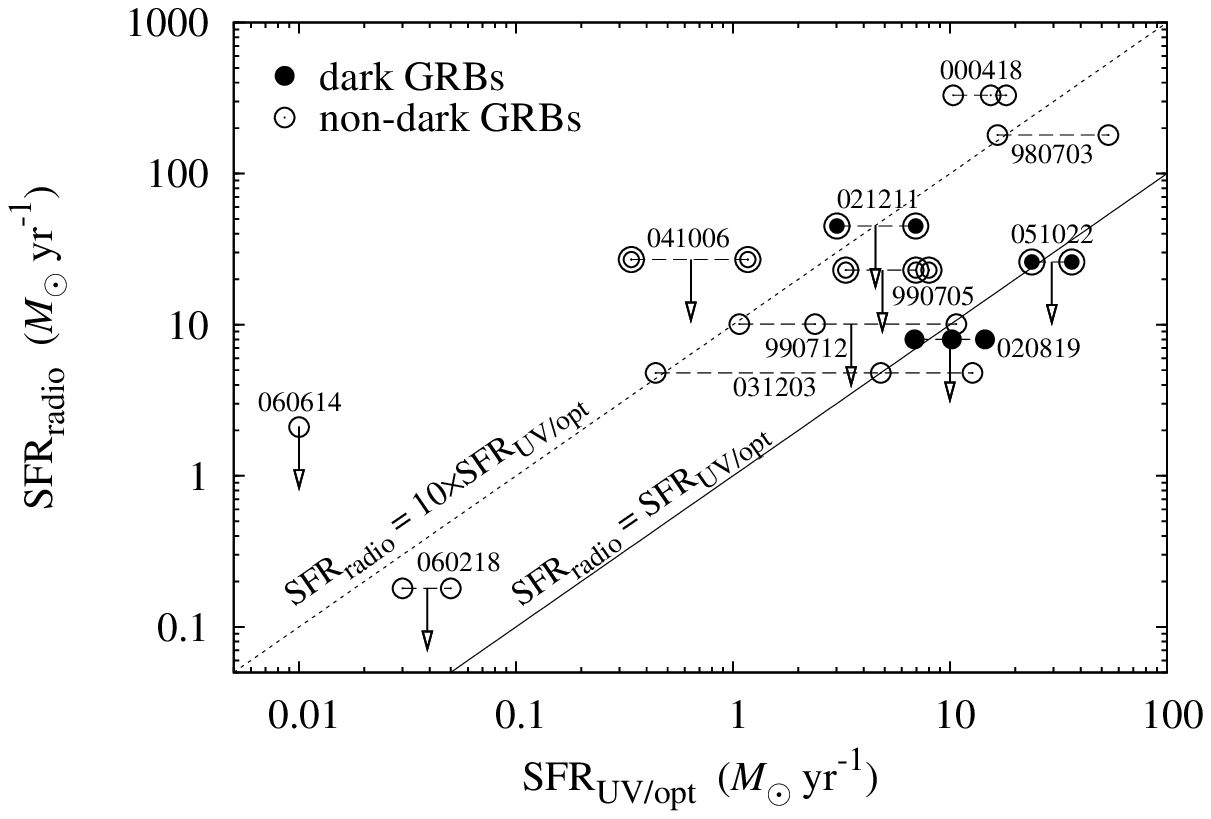}
\caption{
Comparison between SFRs derived from radio observations and extinction-corrected SFRs derived from UV/optical observations. 
We plot the results of this work and previous works: 
Radio data are obtained from \cite{berg03} and \cite{stan10} and UV/optical data are obtained from \cite{goro03, sava09, sven10, leve10a, leve10b}. 
Host galaxies of this work are encircled. 
Filled and open symbols denote dark and non-dark GRBs, respectively. 
Diagonal solid and dotted lines indicate the radio SFRs equal to and 10 times the UV/optical SFRs, respectively. 
Horizontal dashed lines represent the GRB host galaxies with different UV/optical SFRs in literature. 
}
\label{fig:sfr}
\end{figure}

In Figure~\ref{fig:sfr}, we compare the SFRs of the GRB host galaxies derived from the radio observations with those derived from the UV/optical observations. 
The data are taken from our results and literature. 
To check whether there is a difference in obscured star formation between dark and non-dark GRB host galaxies, we differentiate the dark GRBs (solid symbols) from the non-dark GRBs (open symbols). 
The plot shows, that in most cases, the GRB host galaxies do not have a large amount of obscured star formation irrespective of whether they are dark or not. 
This indicates that the dark GRB host galaxies are not always dusty star-forming galaxies \citep[e.g.,][]{lefl06}. 
Recent deep, systematic follow-up observations of GRBs at the optical/NIR region suggest that a significant fraction of GRBs take place in dust-obscured regions \citep[e.g.,][]{perl09, grei11}. 
It is possible that the sample selection could affect the properties of the GRB host galaxies. 
We collect the host galaxies studied with UV/optical wavelengths, which are biased against dusty galaxies. 
The sample size studied here is still small and more observations with extinction-free methods are needed. 
Another possible reason is that the GRB host galaxies studied here are all at $z \lsim 1$, where low-luminosity galaxies account for the large proportion of star-forming activity \citep{lefl05}. 
Recent IR observations uncovered that star formation at $z \gsim 1$ is dominated by IR-luminous dusty galaxies \citep[e.g.,][]{lefl05, take05} and the fraction of dusty GRB host galaxies might increase at higher-redshift.

\subsection{Cause of Dark GRBs}

There is a debate regarding the cause of dark GRBs. 
The possible explanations are that they are 
(1) obscured by dust in host galaxies, 
(2) intrinsically faint, 
and (3) at high redshifts. 
We do not need to consider the possibility of high-redshift GRBs for our sample because they are at $z \le 1$. 
Given the null results of our deep radio observations, it is not likely that the dark GRB host galaxies in our sample are globally obscured by dust.

The optical afterglow spectrum of GRB~021211 shows that the extinction in the line of sight of the GRB is negligible \citep{holl04}. 
The observed optical afterglow of GRB~021211 is thought to be intrinsically faint owing to the fast fading afterglow \citep[e.g.,][]{fox03, crew03, li03}; 
our results support this scenario.

Our results suggest that the host galaxy of GRB~051022 is not likely a dusty star-forming galaxy. 
The X-ray afterglow spectrum suggests a high column density of neutral hydrogen \citep{naka06, cast07}. 
\cite{cast07} state that the extinction in the host galaxy of GRB~051022 ($A_V = 1.0$) derived from its SED cannot account for the afterglow properties, assuming a Galactic dust-to-gas ratio. 
A possible explanation for the cause of dark GRB~051022 is that the burst occurred inside a giant molecular cloud, where a high column density is expected \citep{cast07, rol07}. 
In some GRB afterglows, the absorption features of H$_2$ and CO molecules along the line of sight of GRBs have actually been observed at the same redshifts of the GRBs \citep{fynb06a, proc09, shef09}. 
Measuring the amount of molecular gas in the GRB host galaxies is important, but, so far, CO emission has not been detected \citep{kohn05, endo07, hats07, hats11}. 
Future deep and high-resolution observations of molecular gas will be required to understand the environments of GRB origins.

\section{Summary}\label{sec:summary}

We performed 16-cm continuum observations of the host galaxies of GRBs~990705, 021211, 041006, and 051022 using the ATCA. 
GRBs 021211 and 051022 are known as dark GRBs. 
We failed to detect radio emission in any of the host galaxies. 
The 2$\sigma$ upper limits on the SFRs of the host galaxies of GRBs~990705, 021211, 041006, and 051022 derived from the radio observations are 23, 45, 27, and 26~$M_{\odot}$~yr$^{-1}$, respectively. 
The upper limits on SFRs are less than about 10 times those derived from the UV/optical observations, suggesting that they have no significant dust-obscured star formation, which cannot be traced at UV/optical wavelengths. 
This indicates that dark GRBs do not necessarily occur in galaxies enshrouded by dust. 
The afterglow spectrum of GRB~021211 shows that the dust extinction in the line of sight to the GRB is negligible, and our results suggests the possibility that GRB~021211 is intrinsically faint. 
Given the high column density observed in the afterglow spectrum of GRB~051022, the likely explanation for the dark GRB is dust extinction in the line of sight of the GRB.

\acknowledgments

We would like to acknowledge Sarah Maddison, Robin Wark, Yiannis Gonidakis, and the ATCA staffs for the help they provided during the observation. 
We thank the referee for helpful comments and suggestions. 
The Australia Telescope Compact Array is part of the Australia Telescope National Facility which is funded by the Commonwealth of Australia for operation as a National Facility managed by CSIRO. 
BH is supported by Research Fellowship for Young Scientists from the Japan Society of the Promotion of Science (JSPS). 
TH and KO are supported by the grant-in-aid for Scientific Research on Priority Area (19047003) from the Ministry of Education, Culture, Sports, Science and Technology (MEXT) of Japan.



\begin{thebibliography}{99}
\bibitem[Berger et al.(2003)]{berg03} Berger, E., Cowie, L.~L., Kulkarni, S.~R., Frail, D.~A., Aussel, H., \& Barger, A.~J.\ 2003, \apj, 588, 99 
\bibitem[Castro Cer{\'o}n et al.(2006)]{cast06} Castro Cer{\'o}n, J.~M., Micha{\l}owski, M.~J., Hjorth, J., Watson, D., Fynbo, J.~P.~U., \& Gorosabel, J.\ 2006, \apjl, 653, L85 
\bibitem[Castro-Tirado et al.(2007)]{cast07} Castro-Tirado, A.~J., Bremer, M., McBreen, S., et al.\ 2007, \aap, 475, 101 
\bibitem[Crew et al.(2003)]{crew03} Crew, G.~B., Lamb, D.~Q., Ricker, G.~R., et al.\ 2003, \apj, 599, 387 
\bibitem[Della Valle et al.(2003)]{dell03} Della Valle, M., Malesani, D., Benetti, S., et al.\ 2003, \aap, 406, L33 
\bibitem[Djorgovski et al.(2001)]{djor01} Djorgovski, S.~G., Frail, D.~A., Kulkarni, S.~R., Bloom, J.~S., Odewahn, S.~C., \& Diercks, A.\ 2001, \apj, 562, 654 
\bibitem[Endo et al.(2007)]{endo07} Endo, A., et al.\ 2007, \apj, 659, 1431 
\bibitem[Fox et al.(2003)]{fox03} Fox, D.~W., Price, P.~A., Soderberg, A.~M., et al.\ 2003, \apjl, 586, L5 
\bibitem[Fruchter et al.(2006)]{fruc06} Fruchter, A.~S., et al.\ 2006, \nat, 441, 463 
\bibitem[Fugazza et al.(2004)]{fuga04} Fugazza, D., Fiore, F., Covino, S., et al.\ 2004, GRB Coordinates Network, 2782, 1 
\bibitem[Fynbo et al.(2001)]{fynb01} Fynbo, J.~U., et al.\ 2001, \aap, 369, 373 
\bibitem[Fynbo et al.(2003)]{fynb03} Fynbo, J.~P.~U., et al.\ 2003, \aap, 406, L63 
\bibitem[Fynbo et al.(2009)]{fynb09} Fynbo, J.~P.~U., et al.\ 2009, \apjs, 185, 526 
\bibitem[Fynbo et al.(2006a)]{fynb06a} Fynbo, J.~P.~U., Starling, R.~L.~C., Ledoux, C., et al.\ 2006a, \aap, 451, L47 
\bibitem[Fynbo et al.(2006b)]{fynb06b} Fynbo, J.~P.~U., Watson, D., Th{\"o}ne, C.~C., et al.\ 2006b, \nat, 444, 1047 
\bibitem[Gorosabel et al.(2003)]{goro03} Gorosabel, J., et al.\ 2003, \aap, 409, 123 
\bibitem[Greiner et al.(2011)]{grei11} Greiner, J., et al.\ 2011, \aap, 526, A30 
\bibitem[Hashimoto et al.(2010)]{hash10} Hashimoto, T., Ohta, K., Aoki, K., et al.\ 2010, \apj, 719, 378 
\bibitem[Hatsukade et al.(2007)]{hats07} Hatsukade, B., et al.\ 2007, \pasj, 59, 67 
\bibitem[Hatsukade et al.(2011)]{hats11} Hatsukade, B., Kohno, K., Endo, A., Nakanishi, K., \& Ohta, K.\ 2011, \apj, 738, 33 
\bibitem[Holland et al.(2004)]{holl04} Holland, S.~T., Bersier, D., Bloom, J.~S., et al.\ 2004, \aj, 128, 1955 
\bibitem[Jakobsson et al.(2004)]{jako04} Jakobsson, P., Hjorth, J., Fynbo, J.~P.~U., Watson, D., Pedersen, K., Bj{\"o}rnsson, G., \& Gorosabel, J.\ 2004, \apjl, 617, L21 
\bibitem[Kohno et al.(2005)]{kohn05} Kohno, K., et al.\ 2005, \pasj, 57, 147 
\bibitem[Le Floc'h et al.(2002)]{lefl02} Le Floc'h, E., Duc, P.-A., Mirabel, I.~F., et al.\ 2002, \apjl, 581, L81 
\bibitem[Le Floc'h et al.(2006)]{lefl06} Le Floc'h, E., Charmandaris, V., Forrest, W.~J., Mirabel, I.~F., Armus, L., \& Devost, D.\ 2006, \apj, 642, 636 
\bibitem[Le Floc'h et al.(2005)]{lefl05} Le Floc'h, E., Papovich, C., Dole, H., et al.\ 2005, \apj, 632, 169 
\bibitem[Levesque et al.(2010a)]{leve10a} Levesque, E.~M., Berger, E., Kewley, L.~J., \& Bagley, M.~M.\ 2010a, \aj, 139, 694 
\bibitem[Levesque et al.(2010b)]{leve10b} Levesque, E.~M., Kewley, L.~J., Graham, J.~F., \& Fruchter, A.~S.\ 2010b, \apjl, 712, L26 
\bibitem[Li et al.(2003)]{li03} Li, W., Filippenko, A.~V., Chornock, R., \& Jha, S.\ 2003, \apjl, 586, L9 
\bibitem[Lilly et al.(1996)]{lill96} Lilly, S.~J., Le Fevre, O., Hammer, F., \& Crampton, D.\ 1996, \apjl, 460, L1 
\bibitem[Madau et al.(1996)]{mada96} Madau, P., Ferguson, H.~C., Dickinson, M.~E., et al.\ 1996, \mnras, 283, 1388 
\bibitem[Nakagawa et al.(2006)]{naka06} Nakagawa, Y.~E., Yoshida, A., Sugita, S., et al.\ 2006, \pasj, 58, L35 
\bibitem[Perley et al.(2009)]{perl09} Perley, D.~A., et al.\ 2009, \aj, 138, 1690 
\bibitem[Prochaska et al.(2009)]{proc09} Prochaska, J.~X., Sheffer, Y., Perley, D.~A., et al.\ 2009, \apjl, 691, L27 
\bibitem[Rol et al.(2007)]{rol07} Rol, E., van der Horst, A., Wiersema, K., et al.\ 2007, \apj, 669, 1098 
\bibitem[Salvaterra et al.(2009)]{salv09} Salvaterra, R., et al.\ 2009, \nat, 461, 1258 
\bibitem[Savaglio et al.(2009)]{sava09} Savaglio, S., Glazebrook, K., \& Le Borgne, D.\ 2009, \apj, 691, 182 
\bibitem[Sault et al.(1995)]{saul95} Sault, R.~J., Teuben, P.~J., \& Wright, M.~C.~H.\ 1995, Astronomical Data Analysis Software and Systems IV, 77, 433 
\bibitem[Sheffer et al.(2009)]{shef09} Sheffer, Y., Prochaska, J.~X., Draine, B.~T., Perley, D.~A., \& Bloom, J.~S.\ 2009, \apjl, 701, L63 
\bibitem[Soderberg et al.(2006)]{sode06} Soderberg, A.~M., Kulkarni, S.~R., Price, P.~A., et al.\ 2006, \apj, 636, 391 
\bibitem[Stanek et al.(2003)]{stan03} Stanek, K.~Z., et al.\ 2003, \apjl, 591, L17 
\bibitem[Stanek et al.(2005)]{stan05} Stanek, K.~Z., Garnavich, P.~M., Nutzman, P.~A., et al.\ 2005, \apjl, 626, L5 
\bibitem[Stanway et al.(2010)]{stan10} Stanway, E.~R., Davies, L.~J.~M., \& Levan, A.~J.\ 2010, \mnras, 409, L74 
\bibitem[Svensson et al.(2010)]{sven10} Svensson, K.~M., Levan, A.~J., Tanvir, N.~R., Fruchter, A.~S., \& Strolger, L.-G.\ 2010, \mnras, 405, 57 
\bibitem[Svensson et al.(2011)]{sven11} Svensson, K.~M., Tanvir, N.~R., Perley, D.~A., et al.\ 2011, arXiv:1109.3167 
\bibitem[Takeuchi et al.(2005)]{take05} Takeuchi, T.~T., Buat, V., \& Burgarella, D.\ 2005, \aap, 440, L17 
\bibitem[Tanvir et al.(2009)]{tanv09} Tanvir, N.~R., et al.\ 2009, \nat, 461, 1254 
\bibitem[Totani(1997)]{tota97} Totani, T.\ 1997, \apjl, 486, L71 
\bibitem[Vreeswijk et al.(2003)]{vree03} Vreeswijk, P., Fruchter, A., Hjorth, J., \& Kouveliotou, C.\ 2003, GRB Coordinates Network, 1785, 1 
\bibitem[Yun \& Carilli(2002)]{yun02} Yun, M.~S., \& Carilli, C.~L.\ 2002, \apj, 568, 88 
\end{thebibliography}
\end{document}